# Cyber-Physical Systems – eine Herausforderung für die Automatisierungstechnik?

Prof. Dr.-Ing. S. Kowalewski, Prof. Dr. B. Rumpe, Dipl.-Ing. A. Stollenwerk, RWTH Aachen

Kurzfassung

Der Beitrag befasst sich mit den methodischen Herausforderungen, die durch die Verbreitung der Cyber-Physical Systems (CPS) in der Automatisierungstechnik entstehen, und stellt Lösungsansätze vor. Nach einer Behandlung des Begriffs CPS werden zunächst die allgemeinen, IT-bezogenen Fragestellungen angesprochen, die gemeinsam mit der Informatik gelöst werden müssen. Danach gehen wir auf die Herausforderungen ein, deren Behandlung spezifisch automatisierungstechnische Kernkompetenzen erfordern und skizzieren für eine beispielhafte Problemstellung, den Umgang mit Änderungen in der physikalischen Umgebung, wie entsprechende Lösungen aussehen können.

Abstract

We discuss challenges to control systems engineering arising from the advent of cyber-physical systems (CPS). After discussing the terminology, general, IT-related issues are treated which need cooperation with computer science, in particular software engineering. Then we study those challenges that require specific core competencies from control systems engineering. We sketch solution approaches for the exemplary problem of dealing with changes in the physical environment of a CPS.

1. Einführung

In wenigen Jahren ist der Begriff der *Cyber-Physical Systems (CPS)* zum dominierenden „Buzzword" im Bereich der von technischen Anwendungen motivierten internationalen IT-Forschung geworden [1]. Ausgangspunkt waren die USA, wo 2007 auf Basis einer Empfehlung des *President's Council of Advisors on Science and Technology (PCAST)* ein Forschungsprogramm der National Science Foundation (NSF) mit dem Titel *Cyber-Physical Systems* eingerichtet wurde, in dem seit 2009 über 65 Projekte im Umfang von fast 60 Millionen US$ gefördert werden. 2010 wurden in einem Folgebericht des PCAST der

weiterhin bestehende Forschungsbedarf für CPS festgestellt und das Programm zunächst bis 2013 und mit einem Budget von über 30 Millionen US$ verlängert.

Mittlerweile hat diese Welle auch Deutschland erreicht. Hier wurde im Auftrag der acatech und mit **Unterstützung des BMBF eine „Forschungsagenda CPS" erarbeitet**, die die wirtschaftliche und gesellschaftliche Bedeutung der CPS hervorhebt [2, 3]. Das oben genannte NSF-Programm unterscheidet sich von der acatech-Agenda in zwei wesentlichen Punkten: Das stark von Industrievertretern geprägte PCAST sprach sich in seiner Empfehlung für eine Förderung der Grundlagenforschung zum Thema CPS gezielt an den US-Universitäten und nicht in der Industrie aus. Und die von der NSF geförderten Themen werden maßgeblich mit geprägt von Wissenschaftlern/-**innen aus der „Control Community",** d.h. der regelungs- und automatisierungstechnischen Forschung, während [2] und [3] von Informatikern verfasst wurde. Vor diesem Hintergrund stellt sich die Frage, welchen Beitrag die Automatisierungstechnik in Deutschland zum Thema CPS leisten kann, wie sie von den grundlegenderen Fragestellungen der CPS profitiert und wie die Schnittstellen zu den Fragestellungen aus der Informatik aussehen.

Der Aufsatz führt eine Analyse der CPS-induzierten methodischen Herausforderungen durch, aus der sich neben allgemeinen IT-Fragestellungen auch Forschungsaufgaben ergeben, deren Lösung einen originär regelungs- oder automatisierungstechnischen Beitrag erfordert. Wir untersuchen außerdem, an welcher Stelle und mit welchen Ansätzen die Kooperation zwischen Automatisierungstechnik und Informatik die CPS-Forschung wesentlich voran bringen kann. Die Aussagen werden durch eigene Arbeiten illustriert.

2. Der Begriff CPS
Der Begriff *Cyber-Physical Systems* wurde von Edward Lee geprägt. Eine seiner frühen Definitionen ist [4]:

> *„Cyber-Physical Systems are integrations of computation with physical processes. Embedded computers and networks monitor and control the physical processes, usually* **with feedback loops where physical processes affect computations and vice versa."**

Diese Charakterisierung ist offensichtlich nicht ausreichend, da danach alle digitalen Regler, Steuerungen und Prozessleitsysteme der letzten knapp 50 Jahre CPS wären. Dass CPS mehr sind als klassische Automatisierungssysteme, wird aber deutlich, wenn man die häufig genannten Beispiele betrachtet, z.B. die dezentrale Erzeugung und Verteilung von Energie (**„Smart Grid"), Fahrerassistenz**- und Verkehrssysteme auf der Basis einer Vernetzung von **Fahrzeugen untereinander und mit ihrer Umgebung („Car**-to-**X"),** vernetzte medizintechnische Systeme mit Kopplung von körpernaher Sensorik und Fernüberwachung (**"eHealth"), oder vert**eilte, flexible, selbstkonfigurierende Produktionsanlagen. Besonders die domänenübergreifende Verbindung solcher Systeme, z.B. Smart Grids mit Elektrofahrzeugen und der städtischen Infrastruktur, erhöht noch einmal die Komplexität. In [2, 3] werden CPS daher als Zwischenstufe der Evolution von *vernetzten eingebetteten Systemen* zum so genannten *„Internet der Dinge, Daten und Dienste"* betrachtet.

Was macht also ein verteiltes Automatisierungssystem zu einem CPS? Aus unserer Sicht lässt sich eine tragfähige Taxonomie auf den folgenden beiden Aspekten aufbauen:

- *Verkopplung über geschlossene vs. offene Netze*
  Unter geschlossenen Netzen werden dabei exklusiv für die Automatisierungsfunktion eingerichtete und betriebene Netze verstanden (z.B. Feldbusse), während offene Netze ursprünglich oder zusätzlich auch anderen Kommunikationsaufgaben dienen (z.B. Internet).
- *Domänenbegrenzte vs. domänenübergreifende Funktionalität*
  Unter Domänen werden dabei Anwendungsbereiche mit kohärenten und abgegrenzten Randbedingungen, z.B. Normen, Fach-Communities, eingespielten Praktiken, Entwicklungskulturen etc., verstanden. Beispiele sind Automotive, Hausautomatisierung, Medizintechnik oder Prozessleittechnik.

Aus der Ausprägung dieser Aspekte ergeben sich die folgenden vier Stufen, die am Beispiel eines Autositzes illustriert werden. (Der **Begriff "Automatisierungssystem"** kann hier auch durch **"eingebettetes System" ersetzt werden**.)

1. Nicht vernetzte Automatisierungssysteme
   Beispiel Autositz: Lokal wirkende Steuergeräte, z.B. zur Positionseinstellung.
2. Domänenbegrenzt geschlossen vernetzte Automatisierungssysteme
   Beispiel Autositz: Kopplung von Positionseinstellung und Fahrerprofil-Management oder der Beifahrererkennung über Gewichtssensoren mit der Airbagsteuerung.

3. Domänenbegrenzt offen vernetzte Automatisierungssysteme
   Beispiel Autositz: Übermittlung der besetzten Sitzplätze über WLAN an die Lichtsignal-**Zugangssteuerung zu „Priority Lanes"** für voll besetzte Fahrzeuge
4. Domänenübergreifend offen vernetzte Automatisierungssysteme
   Beispiel Autositz: Kopplung der Überwachung von Vitalfunktionen des Fahrers durch Sensorik im Sitz über Internet mit eHealth-Systemen oder der Notfallambulanz

Eine domänenübergreifende geschlossene Vernetzung ist nicht zu erwarten. In der Literatur finden sich CPS-Beispiele aus den Stufen 2 bis 4. Aus unserer Sicht sollte der Begriff aber auf die beiden höchsten Stufen begrenzt bleiben, da Systeme der Stufe 2 (und zum Teil der Stufe 3) schon existierten, bevor der Begriff CPS eingeführt wurde.

Häufig wird auch die Fähigkeit, mit Änderungen umzugehen, z.B. durch Adaption, Selbstkonfiguration o.ä., als Merkmal von CPS genannt. Auf der Basis der oben eingeführten Taxonomie ergibt sich die Notwendigkeit für CPS, autonom auf Änderungen ihrer Umgebung zu reagieren, durch die Tatsache, dass domänenübergreifend offen vernetzte Systeme nicht als Ganzes entworfen und in ungeänderter Form auf Dauer betrieben werden können. Änderungen sind inhärenter Bestandteil des Lebenszyklus solcher Systeme und stellen einen Aspekt der Herausforderungen dar, die in den nächsten beiden Abschnitten behandelt werden.

3. Gemeinsame Herausforderungen und Lösungsansätze mit der Informatik

Aus der Charakterisierung der CPS wird unmittelbar deutlich, dass sie erhebliche Herausforderungen an die sie entwickelnden und betreibenden Disziplinen stellen. Beispiele sind die komplexe Interaktion von physikalischer Anlage, steuernder Software und Kommunikationsnetzen; heterogene Systemstrukturen mit inkompatiblen, sich verändernden Schnittstellen; oder zur Entwurfszeit nicht vorhersehbare Änderungen im Betrieb des CPS (z.B. nachgeladene Applikationen, neue Anforderungen, neuer Systemkontext). Ein Teil dieser Herausforderungen ergibt sich aus der allgemeinen Komplexität z.B. durch Größe, Heterogenität, Verteiltheit und Strukturdynamik der Systeme. Zu ihrer Bewältigung ist eine gemeinsame Anstrengung mit der Informatik notwendig. Wir beleuchten daher in diesem Abschnitt diese Art von Herausforderungen und passende Lösungsansätze aus der Sicht der Informatik. Im nächsten Abschnitt gehen wir dann auf die spezifisch automatisierungs-technischen Aspekte ein.

Die Behandlung von CPS wird in der Informatik dazu führen, dass die beiden bisher relativ getrennt operierenden Disziplinen des Systems Engineering und des Software Engineering stärker verzahnt agieren müssen und werden. Software Engineering stellt seit längerem einen Baukasten an Konzepten, Techniken und Methoden für die effiziente Entwicklung von Software zur Verfügung und verfeinert diesen permanent. Dem Software Engineering kommt dabei oft die Rolle des Werkzeug- und Methodenanbieters sowie des Integrators zu, um die in anderen Disziplinen der Informatik entwickelten Komponenten anzupassen und zu integrieren. Als Beispiele solcher Komponenten seien neuartige Protokolle zur effizienteren Übertragung in unsicheren Netzen oder zur deterministisch vorhersagbaren Kommunikation, Verfahren zur Bildgebung, zur Gestensteuerung, zur Sensordatenfusion etc. genannt.

Eine wesentliche Charakteristik von CPS ist die *Komplexität*. Das Portfolio des Software Engineering bietet hierfür gemäß dem bereits in der Antike **bekannten Motto „Teile und herrsche" eine Reihe von** Methoden und Methodiken, die die komplexe Aufgabe der ganzheitlichen Systemerstellung, -betreuung und –wartung in einzelne Aktivitäten zerlegen und diese gemäß einem geeigneten Vorgehensmodell in Iterationen verzahnen. Es ist zu erwarten, dass insbesondere agile Methoden, wie Extreme Programming [5] und Scrum [6], geeignet sind, adäquat mit Änderungen und Unsicherheiten in der lebenslangen Entwicklung von CPS umgehen zu können. Diese Methoden sind meist flexibel organisiert und können um neue Aktivitäten zur Systementwicklung ergänzt werden.

Ausgelöst einerseits durch die intensive Vernetzung und damit permanente Verbindung zu den Entwicklern bzw. dem „AppStore" und andererseits durch die auch in eingebetteten Systemen dringender notwendige Aktualisierung von Sicherheitsupdates wird die klassische Trennung zwischen Projekt und Betrieb immer weiter verschwimmen. Zumindest softwareseitige permanente Updates werden die Regel und lassen sich insbesondere in der Automatisierungstechnik auch für Optimierungen nutzen. Die traditionellen Projekt-Vorgehensmodelle werden daher um Methodiken wie Refactoring [7] ergänzt, die es erlauben das System permanent zu adaptieren und zu erweitern. Umbau durch Refactoring kann auf Modellen des Systems gut geplant werden und dürfte sich in beschränkter Form auch für Hardware durchsetzen, wenn diese flexible Umschaltungen oder kostengünstige Umbauten erlaubt und ist auch für hochkritische Systeme möglich [8].

Der Einsatz von Refactoring erfordert ein Umdenken bei der Qualitätssicherung. Speziell die Nutzung von *Simulationstechniken* [9] zur Entwicklung und Anpassung von komplexer Steuerungssoftware ist interessant, weil so komplexeste Situationen virtuell und damit schnell und kostengünstig nachgebaut werden können, die ansonsten nicht adäquat getestet werden könnten. *Virtuelle Systementwicklung* wird ein wichtiger Baustein zukünftiger Methoden.

Virtuelle Entwicklungsprojekte werden massiv auf der Nutzung von *Modellierungstechniken* beruhen. Modelle können genutzt werden, Anforderungen an das Gesamt-System aus verschiedenen Sichten zu beschreiben, die Architektur des Systems aus physikalischer, elektrischer, softwaretechnischer, hydraulischer etc. Sicht zu definieren oder den Systemkontext zu modellieren, der bei der virtuellen Absicherung eine große Rolle spielt. Als Modellierungstechniken stehen hierfür die UML [10], die SysML [11] oder domänenspezifische Sprachen [12] verschiedenster Ausprägungen zur Verfügung. Ihre Verzahnung mit modellbasierten Werkzeugen aus der Regelungs- und Automatisierungstechnik wie Matlab/Simulink oder Modelica ist aber noch wenig ausgeprägt. Zwar bietet die in der Softwareentwicklung sehr verbreitete UML auch gewisse Modellierungskonzepte für eingebettete Systeme an, leidet aber unter nicht ausreichender Standardisierung zur Verfügung stehender Werkzeuge. Eine besondere Herausforderung wird die domänenübergreifende Integration der *Modellierungswerkzeuge sein*, die von der physischen und elektrischen Struktur über Softwaresteuerung, Datenhaltung, Diagnose bis hin zur Einbindung der CPS in die Geschäftsmodellierung idealerweise alles integriert. Damit kann zum Beispiel die Adaption der Geschäftsziele direkt bis zum Umbau der Anlagenstruktur und zur Anpassung der Softwaresteuerung verfolgt und geplant werden und damit dem agilen Unternehmen einen entscheidenden Wettbewerbsvorteil geben.

Die steigende *Intelligenz* („Smartness") des **Automatisierungss**ystems erfordert eine deutlich bessere Einbeziehung der Systemumgebung in Ablauftests und Absicherung. Diese Intelligenz wird aus zwei wesentlichen Gründen für CPS möglich und eine Herausforderung: Zum einen werden in Zukunft große Mengen an Sensoren verschiedenster Arten zur Verfügung stehen, die die Systemumgebung erkennen. Die verarbeitende Software kann diese dann verstehen und darauf reagieren. Zum anderen besteht durch die permanente Verbindung ins Internet bzw. zu einer rechenintensiven und Datenmengen verarbeitenden *Cloud* die Möglichkeit, wesentlich mehr Information in die Entscheidungsprozesse eines CPS (mit oder ohne Mensch) einfließen zu lassen. Die Cloud kann natürlich auch zur Entwicklung und zur Betriebszeit genutzt werden, um emergentes Verhalten, Verhalten bei Betriebsstörungen, Betrieb bei adaptierter Software oder umgebauten Anlagen vorherzusagen. Solche simulativen Vorhersagen werden aber auch für den optimierten Betrieb im Regelfall als auch für schnelle und adäquate Kontrolle im Störfall wichtiger.

Für die *Sicherheit und Zuverlässigkeit* adaptiver, sicherheitsrelevanter Systeme insbesondere auch gegen Attacken von außen bietet die Informatik ein ausgefeiltes Portfolio an Maßnahmen, das von Firewalls über spezielle Programmierrichtlinien sicherheitsrelevanter Software, redundanten Ausfallkonzepten, dynamischer Selbstüberwachung, Sandboxing evtl. kritischer Software bis hin zur Verifikation von Software reicht. Für adaptive Systeme, bei denen während des Betriebs Software nachgeladen wird, kann als besondere Sicherheit zum Beispiel das Verfahren eingesetzt werden, das nur restringierte und damit automatisch prüfbare Modelle einer geeigneten DSL geladen werden oder der Code seinen eigenen Nachweis zur Korrektheit mitbringt, der zur Ladezeit in der Anlage geprüft wird [13]. Zur Erstellung solcher Nachweise können zum Beispiel Model-Checking-Techniken [14] eingesetzt werden. Es ist auch davon auszugehen, dass die Analyse hybrider dynamischer Systeme [15] im Zusammenhang mit CPS eine größere Relevanz gewinnen wird.

Die Informatik und als integrativer Teil das Software Engineering bieten also ein Portfolio an Techniken, die es zu nutzen gilt, um die Herausforderungen durch CPS auch in der Automatisierungstechnik zu bewältigen. Die Techniken müssen auf CPS weiter adaptiert und teilweise auch substantiell durch neue Techniken ergänzt, insbesondere aber neu aufeinander abgestimmt werden, um weiterhin effektiv einsetzbar zu sein. Die aktuell anstehende Welle von CPS-orientierten Projekten wird diese in den nächsten zehn bis fünfzehn Jahren entwickeln und in die Praxis bringen. Jede der genannten Domänen, zu denen auch die Automatisierungstechnik zählt, wird wird von CPS-Ansätzen profitieren und ihre eigenen domänenspezifischen Lösungen bzw. Adaptionen generischer Lösungen benötigen, um die Möglichkeiten der neuen Techniken des Engineerings von CPS voll auszuschöpfen.

## 4. Besondere Herausforderungen an die Automatisierungstechnik und Lösungsansätze

Da die Interaktion mit physikalischen Prozessen in geschlossenen Wirkungskreisen ein definitorischer Aspekt von CPS ist, gehen die Herausforderungen an die Entwurfs- und

Analysemethoden zum Teil auch über die Domäne der Informatik hinaus und betreffen Kernkompetenzen der Automatisierungstechnik. Hier ist an erster Stelle die Beherrschung der Systemdynamik in den entstehenden Regelkreisen zu nennen. Insbesondere das Schließen der Kreise über nichtdeterministische Kommunikationsnetze stellt neue Anforderungen an Reglerentwurfs- und Systemanalysemethoden. Gleiches gilt für die Notwendigkeit von Umschaltungen durch Strukturänderungen. Zu beiden Fragestellungen hat die Regelungstechnik in den letzten Jahren unter den Stichworten *Networked Control* [16, 17], *Hybrid Control* [15] und *Reconfigurable Control* [18] u.a. in DFG-Schwerpunktprogrammen passende Methoden erarbeitet.

Ebenfalls spezifisch automatisierungstechnisch ist die Frage, wie die offene und gar domänenübergreifende Vernetzung von Automatisierungssystemen in der Produktion beherrschbar bleibt und gestaltet werden kann. Diese wird zu Verknüpfungen von Prozessen und Daten über Unternehmen hinweg führen [19, 20]. Entscheidende Fragen werden hier z.B. sein, wie eine übergreifende Produktionssteuerung realisiert werden kann und wie die Migration der existierenden Leittechnik mit Anbindung an mehrere ERP-Systeme aussehen wird.

Schließlich müssen Konzepte bereitstehen, um Änderungen der physikalischen Umgebung **(also der „Strecke") schnell und zuverlässig berücksichtigen zu können.** Neben der klassischen adaptiven Regelung betrifft das auch strukturelle Änderungen wie das Austauschen von Sensoren oder Aktuatoren, die mit möglichst geringem Anpassungsaufwand durchgeführt werden müssen.

Eine besondere Ausprägung dieser Änderungen ist der Wechsel von der modellierten Strecke zur realen Strecke im Rahmen der modellbasierten Entwicklung mit den Schritten Model-In-The-Loop (MIL), Hardware-In-The-Loop (HIL) und Test am realen System [21]. Der Schritt von der simulierten, virtuellen zur realen Strecke sollte möglichst geringe Adaptionen des Reglermodells erfordern. Ein Ansatz, wie dies mit Hilfe von Abstraktionen erreicht werden kann, wird in [22] für den Entwurf von Motorsteuerungssoftware vorgestellt. Bild 1 zeigt die dazugehörige Struktur.

In der Modellierungsphase wird für die Sensoren und Aktuatoren eine Verhaltensmodellierung, einschließlich physikalischer Eigenschaften wie z. B. Auflösungsvermögen, zeitliche Verzögerungen oder Grenzen, hinterlegt, die beim MIL-Test genutzt werden kann.

Bei der Transition auf das reale System müssen dann nur die simulierten Sensoren und Aktuatoren durch die Treiberaufrufe der realen Sensoren ersetzt werden. Eine weitere Adaption der entwickelten Software ist nicht notwendig.

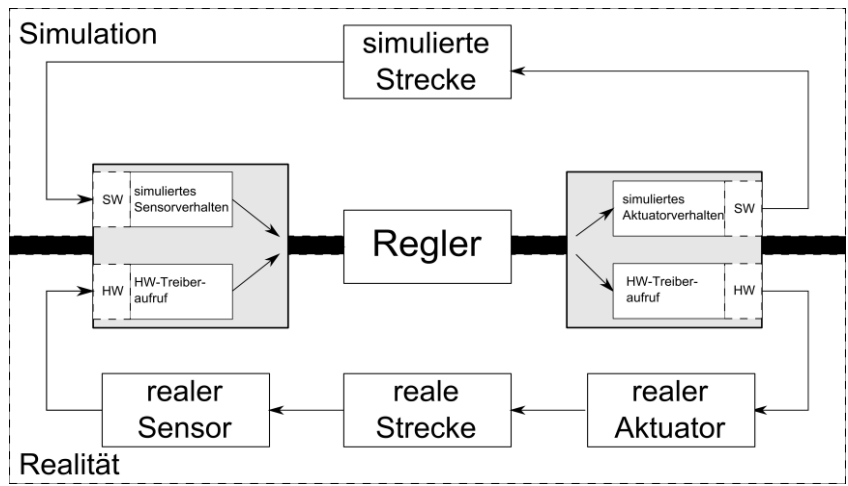

Bild 1:   Struktur zur Umschaltung zwischen simulierter und realer Strecke [22]

Ein zweites Beispiel mit ähnlichem Ansatz ist eine im Rahmen einer intensivmedizinischen Anwendung entwickelte Hardwareplattform, die in einem dezentralen Sicherheitskonzept Anwendung findet [23, 24]. Auf dieser Hardwareplattform werden zum einen Betriebsfunktionen, wie das Auslesen medizinischer Monitore und das Ansteuern der genutzten Aktuatoren, aber auch Sicherheitsfunktionen modellbasiert abgebildet, wie z.B. die Überwachung eines Sensors auf Plausibilität seiner Messwerte bzw. die physikalische Realisierbarkeit von Stellwerten [25, 26].

Um die einzelnen Modelle voneinander trennen zu können und modular austauschbar zu halten, wurde wie in dem vorhergehenden Beispiel eine auf Abstraktionsschichten basierte Softwarearchitektur eingeführt. Basierend auf einem eingebetteten Echtzeitbetriebssystem sind die Hardwareplattform an einen Kommunikationsbus zum Datenaustausch und an den jeweiligen Sensor bzw. Aktuator angebunden. Die auszuführenden Modelle werden über Schnittstellen in einer spezielle Einbettungsschicht (Wrapper) eingebettet. Eine spezielle Datenhaltungsschicht wickelt die Verarbeitung der erzeugten und benötigten Daten ab [26]. Diese Interaktion der verschiedenen Schichten ist schematisch in Bild 2 gezeigt.

Der Wrapper hält initial alle im System vorkommenden Daten vor (V1 bis V6). Die verschiedenen eingebetteten Modelle signalisieren dem Wrapper zur Compilierzeit, welche

Daten benötigt bzw. erzeugt werden, sodass schon zur Compilierzeit nicht benötigte Daten (im Beispiel V4) verworfen werden können. Die verbleibenden Daten werden entsprechend der signalisierten Bedürfnisse an die Datenhaltungsschicht angebunden. Da die Datenhaltungsschicht und die Einbettungsschicht keiner Interaktion des Entwicklers bedürfen, können diese vollständig automatisch generiert werden. Die vorgestellte Softwarearchitektur ermöglicht es außerdem, zur Compilierzeit Aussagen zum Speicherverbrauch und zur maximalen Laufzeit eines Modells zu tätigen. So lassen sich Speicherüberläufe bzw. Verletzungen der Echtzeitanforderungen an ein Modell wirkungsvoll vermeiden.

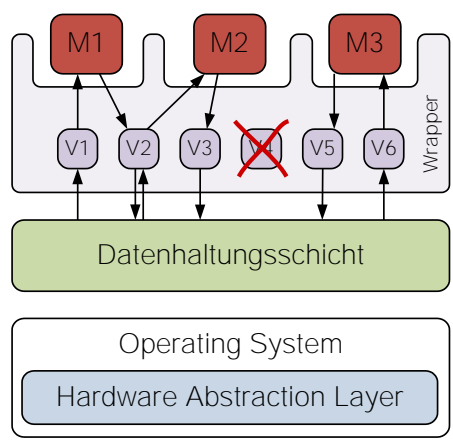

Bild 2: Interaktion der Abstraktionsschichten und Adaption der Softwarearchitektur [26]

Zusammenfassung

CPS unterscheiden sich von klassischen Automatisierungssystemen durch eine Vernetzung, die über nicht für die Automatisierungsfunktion eingerichtete und betriebene Verbindungen (z.B. Internet) realisiert wird, und durch die Möglichkeit domänenübergreifender Funktionalität. Die Tatsache, dass CPS nicht als Ganzes zu entwerfen und zu warten sind, führt zu Komplexität durch Heterogenität und Strukturänderungen und zur Notwendigkeit adaptiven Verhaltens. Adäquate neue Entwurfs- und Analysemethoden können zum Teil auf der Basis von Informatik-Ansätzen erarbeitet werden. Zum Teil erfordern sie aber spezifisch automatisierungstechnische Kompetenzen, insbesondere im Umgang mit der Streckendynamik. Die Automatisierungstechnik ist daher gut beraten, CPS nicht nur als Herausforderung sondern auch als Chance zur Weiterentwicklung ihrer Kompetenz, ihres Methodenportfolios und ihres Geschäftsfeldes zu betrachten.